\documentclass[12pt,epsf]{article}
\usepackage{epsfig}

\newcommand{\setfigure}[2]{\begin{figure}[htbp]
\begin{center}\leavevmode\epsfxsize=5in\epsfbox{#1.eps}\end{center}\caption{#2\label{#1}}
\end{figure}}

\renewcommand{\thanks}[1]{\footnote{#1}} % Use this for footnotes

\newcommand{\be}{\begin{equation}}
\newcommand{\ee}{\end{equation}}
\newcommand{\bea}{\begin{eqnarray}}
\newcommand{\eea}{\end{eqnarray}}

\begin{document}

\pagestyle{empty}

\bigskip\bigskip
\begin{center}
{\bf \large Construction of a Penrose Diagram for an Accreting Black Hole}
\end{center}

\begin{center}
Beth A. Brown\\
NASA Goddard Space Flight Center,
Greenbelt, Md.   20771\\
\bigskip
James Lindesay\footnote{e-mail address, jlslac@slac.stanford.edu} \\
Computational Physics Laboratory,
Howard University,
Washington, D.C. 20059 
\end{center}
\bigskip

\begin{center}
{\bf Abstract}
\end{center}
A Penrose diagram is constructed for a spatially coherent
black hole that accretes at stepwise steady rates as measured
by a distant observer from an initial state described
by a metric of Minkowski form.  Coordinate lines
are computationally derived, and radial light-like
trajectories verify the viability of the diagram. 
Coordinate dependencies of significant features, such as the
horizon and radial mass scale, are clearly
demonstrated on the diagram. 
The onset of a singularity at the origin is shown to
open a new region in space-time that contains the interior
of the black hole.
\bigskip \bigskip \bigskip

\setcounter{equation}{0}
\section{Introduction}
\indent

Astrophysical black holes are expected to evolve, i.e. accrete
or evaporate, over time. 
One expects that the initiation and final
evaporation processes of black hole dynamics
should involve spatial coherence on scales comparable
to those describing the medium scale structure of the
space-time, since the geometries during those periods are
defined by micro-physics.  It is of particular interest to examine
the correspondence of low-curvature space-time with a
dynamic black hole geometry.

The non-orthogonal temporal coordinate associated with the
river model of black holes\cite{rivermodel, Nielsen} has been
shown to provide a convenient parameter for describing the
evolution of a black hole without physical singularities 
in the vicinity of the horizon.  In a previous paper, we
examined an example dynamic black hole undergoing a
steady rate of evaporation\cite{BABJL1}.  To complement
evaporation, in this paper we
will examine the global causal structure of a black hole
undergoing periods of stepwise steady accretion. 
This is accomplished via the construction of a computationally
derived Penrose diagram.  The viability of this diagram can be
verified by computing and plotting radial light-like trajectories
as curves of slope $\pm$unity. 
The complete
evolutionary cycle of a black hole should involve a merger
of the techniques involved in these companion papers\cite{JLNSBP08}.

\setcounter{equation}{0}
\section{Form of the Metric and Conformal Coordinates
\label{Section2}}

\subsection{Form of the metric}
\indent

The dynamic
space-time metric will be assumed to take the form
\be
ds^2 = -\left (1-{R_M (ct) \over r} \right ) (dct) ^2 +
2 \sqrt{{R_M (ct) \over r}} dct \, \, dr + dr^2 + r^2 \, 
(d\theta^2 + sin^2 \theta \, d\phi^2).
\label{metric}
\ee
In this equation, $R_M (ct) \equiv 2 G_N M(ct) / c^2$ is a time-dependent
form of the Schwarzschild radius that we refer to as the
\emph{radial mass scale}.   This metric was examined as a dynamic
extension of the river model of (static) black holes discussed in the
literature\cite{rivermodel, JLMay07}.  The metric takes the
form of a Minkowski space-time
both asymptotically ($r \rightarrow \infty $) as well as when the
radial mass scale vanishes ($R_M(ct) \rightarrow 0$).  Therefore,
the temporal and radial coordinates are those of an observer
far from the black hole.  All curvature components generated
by this metric are non-singular near the light-like surface defining
the horizon, $\dot{R}_H = 1 - \sqrt{R_M \over R_H}$,
as well as the surface defined by the radial mass
scale, $R_M(ct)$.  This simplifies descriptions of the physics
in the vicinity of $R_H$ and $R_M$.

\subsection{Form of conformal coordinates}
\indent

The form of the conformal temporal and radial coordinates
that are used for the construction of the Penrose diagram
have been developed in companion papers\cite{JLMay07, BABJL1}:
\be
\begin{array}{c}
ct_* = {r \over 2} \left (
 exp \left [ \int ^ {R_M (ct) \over r} { \left (
1 + \sqrt{\zeta '}  \right ) d \zeta '  \over
\left \{ \zeta' \left ( 1 + \sqrt{\zeta'} \right )
+ \dot{R}_M \right \} }
\right ]  -  
 exp \left [ \int ^ {R_M (ct) \over r} { \left (
1 - \sqrt{\zeta '}  \right ) d \zeta '  \over
\left \{ \zeta' \left ( 1 - \sqrt{\zeta'} \right )
- \dot{R}_M \right \} }
\right ] \right ) \\ \\
r_* = {r \over 2} \left (
 exp \left [ \int ^ {R_M (ct) \over r} { \left (
1 + \sqrt{\zeta '}  \right ) d \zeta '  \over
\left \{ \zeta' \left ( 1 + \sqrt{\zeta'} \right )
+ \dot{R}_M \right \} }
\right ]  + 
 exp \left [ \int ^ {R_M (ct) \over r} { \left (
1 - \sqrt{\zeta '}  \right ) d \zeta '  \over
\left \{ \zeta' \left ( 1 - \sqrt{\zeta'} \right )
- \dot{R}_M \right \} }
\right ] \right )  .
\end{array}
\label{conformal}
\ee
The transformations are valid for constant accretion
rates $\ddot{R}_M=0$. 
These equations relate the space-time coordinates of an
asymptotic observer $(ct, r)$ with the conformal coordinates
$(ct_* , r_*)$.  In subsequent calculations, the constants
associated with the integrations are chosen so that the
conformal space-time parameters $(ct_*,r_*)$
asymptotically (i.e., as ${R_M(ct) \over r} \rightarrow 0$)
correspond to Minkowski
space-time parameters $(ct,r)$, which are likewise conformal.

\setcounter{equation}{0}
\section{Penrose Diagram of the Accreting Black Hole
\label{Section3}}

\subsection{Procedure for constructing the Penrose diagram}
\indent

The diagram that is developed uses hyperbolic tangents of a
scaled multiple of the conformal coordinates in Eq. \ref{conformal} 
to map the infinite domain of those conformal coordinates onto a
finite region.  
More specifically, the vertical coordinate $Y_{t*}$ takes the
form $(tanh({ct_* + r_* \over scale}) + tanh({ct_* - r_*
\over scale}) ) / 2$ and the horizontal coordinate $Y_{r*}$ takes the
form $(tanh({ct_* + r_* \over scale}) - tanh({ct_* - r_*
\over scale}) )/ 2$ in the exterior region. 
In the interior region $r<R_H (ct)$, the time-like and space-like
coordinates ($Y_{t*}$ and $Y_{r*}$)
are interchanged (because of a sign change in the
coefficient of $[-(dct_*)^2 + (dr_*)^2]$ in the
conformal metric as one crosses the
horizon) and shifted in a manner that preserves
the coordinates of the horizon.  The overall
transformation assures that the slopes of
\emph{any} outgoing/ingoing
light-like radial trajectories on the diagram will be $\pm 1$, and
that the diagram
has its domain and range bounded by $\pm 1$. 

The initial metric will be assumed to be of the form of
Minkowski space-time for $t \leq 0 $.  The accretion
begins at $t=0$ from an initial radial mass scale
$R_M(0)=0$, which smoothly transitions the dynamic metric
in Eq. \ref{metric} from a Minkowski metric form.  
The radial coordinates must smoothly match across the
volume $t=0$, since $4 \pi r^2$ measures the area
of any sphere of radial coordinate $r$ in both metric forms. 
Since the conformal coordinates developed require a steady rate
of accretion, changes in that rate (if desired) can be modeled
using stepwise steady accretions, and matching
the coordinates of the radial mass
scale $R_M$ (a physical parameter directly related to the
mass of the black hole) across differing rates
with instantaneously matching metrics.

\subsection{Features of the Penrose diagram}
\indent

The Penrose diagram in Figure \ref{AccrPenr} demonstrates
the expected global structure of this spherically symmetric black hole that
accretes at a steady rate of change in the radial mass
scale $R_M (ct)$ with respect to the
distant observer's time coordinate $ct$.  
\setfigure{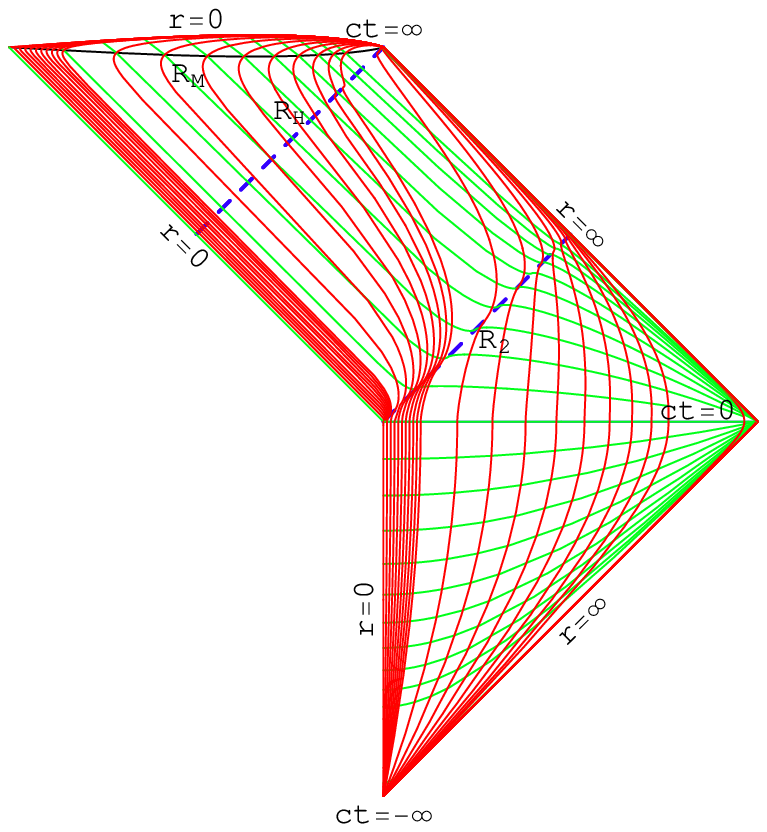}{Penrose diagram for a black hole that accretes steadily
from zero mass at $ct=0$.  Red curves (running vertically in the 
exterior right
hand region) represent curves of constant $r$.  The green curves
represent curves of constant $ct$.  The dashed blue line labeled
$R_H$ represents the horizon.}
In the diagram, the red curves that are time-like in the right hand
regions represent curves of constant $r$, originally graded from
$r=0$ in hundredths, tenths, then in multiples, and decades of the chosen scale. 
The curves of constant radial coordinate $r$ all originate at the bottom corner of the
diagram representing $t=-\infty$, and terminate at the uppermost corner
representing $t=+\infty $.  The green curves that are space-like in the
right hand regions represent curves of constant $ct$ graded in multiples
of the given scale.  All constant $ct$ curves originate on the curve $r=0$
and terminate at the far right corner of the diagram representing $r=\infty $. 
The various $ct=constant$ and $r=constant$ curves each intersect at only
one point on the diagram.  The light-like bounding curves $r=\infty$
on the right are those of a Minkowski space-time.

Prior to the formation of the black hole, the space-time
is taken to be
that described by Minkowski, represented by the lower portion
of the diagram.  At $t=0$, the singularity develops via
an ingoing light-like transition $(ct=0,r=0)$,
during which a new region in the space-time
opens up.  This new region, which contains the interior
of the black hole, is represented by the upper
left quadrant in Figure \ref{AccrPenr}.  The curve
$r=0$ is initially a time-like trajectory bounding the Minkowski
space-time on the left. It then undergoes the light-like transition
to become a space-like trajectory bounding the upper
interior region of the black hole from above.  The Penrose
coordinates of the singularity $r=0$ somewhat matches 
the behavior in the
vertical Penrose coordinate of a static Schwarzschild singularity
$Y_{singularity \,*}^{Schwarzschild}=+1$.  For an accreting
black hole, the radial mass scale $R_M (ct)$ 
(indicated by
the solid black space-like curve between the singularity and
the horizon $R_H$)
can be shown
generally to lie within the horizon\cite{JLMay07}, and it is seen
to mirror the behavior of the singularity $r=0$.  Radial coordinate
curves are seen to transition from time-like to space-like as
they cross the radial mass scale. 
This occurs because outgoing light-like trajectories, which
on this geometry
satisfy $\dot{r}_\gamma=1-\sqrt{R_M / r_\gamma}$,
are temporarily stationary in the radial
coordinate $\dot{r}_\gamma = 0$ at the radial mass scale.

The mass scale and horizon of the black hole are both dynamic. 
The horizon is represented
by the dashed diagonal blue line indicated by $R_H$
in Figure \ref{AccrPenr}, and it
lies completely within the new region of
space-time opened up by the formation of the black
hole singularity. 
Unlike as is the case for a static Schwarzschild black hole, the
horizon $R_H (ct)$ is \emph{not} a $t=\infty $ surface.  Both radial
and temporal coordinate curves are seen to cross the horizon
and radial mass scale on the diagram.  Thus,
the distant observer's time is qualitatively different from
that of a distant Schwarzschild observer, even for a very slowly
accreting black hole.

Finally, there is a region of significant coordinate
distortion along the light-like curve labeled $R_2$,
which corresponds to one of the singular curves of
the conformal metric\cite{JLNSBP08}.  The coefficient of the
factor $[-(dct_*)^2+(dr_*)^2]$ in the metric becomes singular
on the outgoing light-like curve $R_2(ct)$ and vanishes on the horizon
$R_H(ct)$.  However, this anomalous
coordinate behavior near $R_2$ does not correspond to any
physical singularity on the geometry (for example, the
Ricci scalar of the geometry, which is of the form
$\mathcal{R}={3 \dot{R}_M \over 2 r^2} \sqrt{r \over R_M}$,
is non-singular near $R_2$).
For the steadily accreting black
hole, in the distant
future the area of the horizon of the black hole is
unbounded ($R_H \rightarrow \infty $
as $t \rightarrow \infty $). 
However, as was the case for the steadily excreting
black hole\cite{BABJL1},
it remains possible to stay external to the black hole,
but not with a fixed radial coordinate $r$.  A radial coordinate
that is a fixed multiple of that of the horizon remains external
if that multiple is greater than unity.

\setcounter{equation}{0}
\section{Conclusions and Discussion}
\indent

The large-scale geometry of a steadily accreting black hole revealed
several aspects of interest. 
The radial mass scale was seen to have a trajectory internal to
the horizon of the black hole as expected from analytical examinations.
The diagrammatic representation of the speed of light formation of
the singularity was unexpected, but necessary, since low
curvature space-times have non-negative conformal radial coordinates.
This light-like formation of the singularity concurrently expands
the large-scale structure of the space-time. 
Both the radial mass scale and the horizon were seen to lie completely
within the region of space-time opened during the formation of the
singularity. 

The observation that the radial mass scale does
not coincide with the horizon demonstrates that the dynamics
described using our temporal coordinate $t$ is qualitatively
different from what would result from a mass scale
dependency upon a Schwarzschild
observer's temporal coordinate, no matter how small the
rate of accretion. 
Parametric descriptions of both the radial mass scale
and the horizon are non-singular in the coordinates $(ct,r)$. 
In addition, all physical curvatures are well defined near
these trajectories.

The asymptotic correspondence of the black hole description
was seen to fix the scales of the
past and future light-like infinities of the black hole to be the
same as those of Minkowski space-time. 
Using the coordinates employed for the spatially coherent
accreting black hole, when the accretion initiates there is
continuity in the metric with that of static Minkowski space, but a
discontinuity in curvature.  The primary concerns of the present
paper has been the construction of the Penrose diagram rather
than the development of a model for the micro-physics describing
the initial stage of accretion, which likely occurs on time scales small
compared to the resolution of the diagram.  Since the metrics
precisely match at the transition,
it is felt that the large scale structure has
been accurately portrayed.

A Penrose diagram representing
the complete life cycle of a model black hole can
likewise be constructed
using stepwise steady rates of accretion/evaporation, and
matching metric forms across the transitions.  A diagram
demonstrating the initiation of accretion
from a low curvature space-time, an end of accretion and
beginning of evaporation, and an end of evaporation as the
black hole mass vanishes, is presently
being examined by the authors, and its form will be
presented in a future manuscript. 
We ultimately want to examine the micro-physics that allows the
initial accretion and final evaporation processes of a black hole.  This
will involve exploring general dynamics in the mass scale
$R_M(ct)$ in a geometry with rates which can smoothly
transition from and to low curvature space-times.

\begin{center}
\textbf{Acknowledgments}
\end{center}
BAB would like to acknowledge the support of the NASA Administrator's
Fellowship Program.  JL must regretfully inform the reader that
his colleague Dr. Beth A. Brown, the co-author, unexpectedly passed during the
preparation of the final draft of this manuscript.

\end{document}